\documentclass[aps,prl,twocolumn,byrevtex]{revtex4}

\let\oldhat\hat
\renewcommand{\hat}[1]{\oldhat{\mathbf{#1}}}
\usepackage{times,mathptm}
\usepackage[dvips]{graphicx,color}
\usepackage{makecell}
\usepackage{booktabs}
\begin{document}
\title{Comment on `Quasi-One-Dimensional Metal-Insulator Transitions in Compound Semiconductor Surfaces'}
\author{Sun-Woo Kim$^1$, Yoon-Gu Kang$^1$, Hyun-Jung Kim$^2$, and Jun-Hyung Cho$^{1*}$}

\affiliation{$^1$Department of Physics and Research Institute for Natural Sciences, Hanyang University, 222 Wangsimno-ro, Seongdong-gu, Seoul, 133-791, Korea  \\
$^2$ Korea Institute for Advanced Study, 85 Hoegiro, Dongdaemun-gu, Seoul 02455, Korea}

\date{\today}


\maketitle

Based on density-functional theory (DFT) calculations within the generalized gradient approximation (GGA), Zhao $et$ $al$.~\cite{zhao} claimed that one-atom-wide metallic structures formed by selectively bonding of H or Li atoms to GaN(10${\overline{1}}$0) and ZnO(10${\overline{1}}$0) undergo the Peierls-type metal-insulator (MI) transitions, leading to a charge-density-wave (CDW) formation with periodic lattice distortion. However, we here demonstrate that such a CDW phase is due to the artifact of the GGA, while the antiferromagnetic (AFM) ground state is predicted by the hybrid DFT calculation and the exact-exchange plus correlation in the random-phase approximation (EX + cRPA).

For the 1D metallic structure composed of the surface metal atoms (Ga or Zn), Zhao $et$ $al$. [1] found that the semiconducting $p$(1${\times}$2) phase with alternately up-and-down displacements is more stable than the metallic $p$(1${\times}$1) phase, and the $p$(2${\times}$2) phase is further stabilized. The relative total energies (${\Delta}E$) of these phases for GaN(10${\overline{1}}$0)-1H are listed in Table I.

For GaN(10${\overline{1}}$0)-1H, Zhao $et$ $al$.~\cite{zhao} showed that the $p$(1${\times}$1) phase has a Fermi surface nesting that drives the CDW with the up-down buckling distortion. The resulting Peierls-type MI transition produced the $p$(1${\times}$2) phase with ${\Delta}E$ = $-$0.51 eV and $E_g$ = 0.74 eV (see Table I). The further lattice relaxation of $p$(2${\times}$2) gives ${\Delta}E$ = $-$0.57 eV and $E_g$ = 0.99 eV. Interestingly, a ferromagnetic (FM) phase appears in metallic $p$(1${\times}$1), but such spin ordering disappears in semiconducting $p$(1${\times}$2) and $p$(2${\times}$2)~\cite{zhao}. This disappearance of magnetic order may be due to the fact that the local density approximation or GGA tends to stabilize artificially delocalized electronic states due to their inherent self-interaction error (SIE)~\cite{cohen}.

We optimize the structure of GaN(10${\overline{1}}$0)-1H using the all-electron FHI-aims~\cite{aims} code with the GGA functional of PBE~\cite{per}. The present results of ${\Delta}E$ and $E_g$ for the nonmagnetic (NM) $p$(1${\times}$2)- and $p$(2${\times}$2)-CDW phases agree well with those reported in the Letter (see Table I). In order to correct the SIE, we use the HSE hybrid functional~\cite{hse} to calculate ${\Delta}E$ and $E_g$ of various phases including the AFM order. Here, we employed a mixing factor of ${\alpha}$ = 0.32 controlling the amount of exact-exchange energy, which predicts well the observed bulk band-gap of 3.51 eV~\cite{Ga}. We find that the HSE calculation stabilizes the magnetic phases with ${\Delta}E$ = $-$0.95, $-$1.10, and $-$1.11 eV for $p$(1${\times}$1)-FM, $p$(1${\times}$2)-AFM, and $p$(2${\times}$2)-AFM, respectively (see Table I). Meanwhile, ${\Delta}E$ of $p$(1${\times}$2)- and $p$(2${\times}$2)-CDW are $-$0.98 and $-$1.04 eV, respectively. Thus, the HSE calculation shows that the AFM phase is energetically more stable than the CDW formation. This result is confirmed by the EX + cRPA calculation~\cite{rpa,rparesult}.

\begin{table}[ht]
\caption{Total energies [in eV per $p$(2${\times}$2) unit cell] of various phases relative to NM $p$(1${\times}$1) and band gaps (in eV). }
\begin{ruledtabular}
\begin{tabular}{lcccccc}
   & \multicolumn{2}{c}{PBE (Ref. ~\cite{zhao})} & \multicolumn{2}{c}{PBE}  & \multicolumn{2}{c}{HSE} \\
       &   ${\Delta}E$ &  $E_{\rm g}$   & ${\Delta}E$ &  $E_{\rm g}$  &  ${\Delta}E$ &  $E_{\rm g}$ \\ \hline
$p$(1${\times}$2)-CDW    & $-$0.51  &  0.74   &  $-$0.51  & 0.71   &  $-$0.98 & 1.68 \\
$p$(2${\times}$2)-CDW    & $-$0.57  &  0.99   &  $-$0.57  & 0.83   &  $-$1.04 & 1.82 \\
$p$(1${\times}$1)-FM     &   $-$    &  metal  &  $-$0.12  &  metal &  $-$0.95 & 1.04 \\
$p$(1${\times}$2)-AFM    &          &         &           &        &  $-$1.10 & 1.85 \\
$p$(2${\times}$2)-AFM    &          &         &           &        &  $-$1.11 & 1.98 \\
\end{tabular}
\end{ruledtabular}
\end{table}

Our GGA- and hybrid-DFT calculations showed the different predictions for the ground state of GaN(10${\overline{1}}$0)-1H. Contrasting with PBE predicting the $p$(2${\times}$2)-CDW phase, HSE predicted the $p$(2${\times}$2)-AFM phase. It is noted that the metallic $p$(1${\times}$1) phase of GaN(10${\overline{1}}$0)-1H has a narrow band width of 1.24 eV at the Fermi level, while that of ZnO(10${\overline{1}}$0)-1H has a wide band width of 4.08 eV~\cite{zhao}. Such more localized dangling-bond electrons in GaN(10${\overline{1}}$0)-1H are likely to favor the AFM order over the CDW formation, contrasting with the case of ZnO(10${\overline{1}}$0)-1H. The future experimental works are stimulated to confirm these theoretical predictions.

\textbf{Acknowledgements} S.-W. Kim, Y.-G. Kang, and H.-J. Kim equally contributed to this work. We thank Prof. X. Ren for his help on EX + cRPA calculations. This work was supported by the National Research Foundation of Korea grant funded by the Korean government (Nos. 2015M3D1A1070639 and 2015R1A2A2A01003248), KISTI supercomputing center through the strategic support program for the supercomputing application research (KSC-2016-C3-0001), and Center for Advanced Computation (CAC) of Korea Institute for Advanced Study. S.W.K. acknowledges support from POSCO TJ Park Foundation.

\noindent $^{*}$ Coresponding authors: chojh@hanyang.ac.kr


\begin{thebibliography}{99}
\bibitem{rparesult} For the EX + cRPA calculation, we used the periodic slab geometry and the electronic states, obtained using the PBE. The EX + cRPA total-energy difference ${\Delta}E_{\rm EX+cRPA}$ between the CDW and AFM phases is 122 meV per 2${\times}$2 unit cell, which is somewhat larger than the corresponding one (68 meV) obtained using the HSE calculation with ${\alpha}$ = 0.32. However, this value of ${\Delta}E_{\rm EX+cRPA}$ should be influenced by either the self-consistent EX + cRPA calculation or a more rigorous approach including the high-order diagrams beyond cRPA.
\end{thebibliography}
\end{document}